\documentclass[12pt]{article}

\usepackage{epsfig}
\usepackage{amsmath,amssymb}
\usepackage{amsthm}
\usepackage{setspace}
\usepackage{amsbsy}
\usepackage{natbib}
\usepackage{cases}
\usepackage{booktabs}
\usepackage{enumitem}
\usepackage{verbatim}
\usepackage{csquotes}

\newtheorem{theorem}{Theorem}
\newtheorem{proposition}{Proposition}
\newtheorem{lemma}{Lemma}
\newtheorem{definition}{Definition}
\newtheorem{property}{Property}

\DeclareMathOperator*{\argmin}{\arg\!\min}

\begin{document}
\renewcommand{\arraystretch}{1.5}
 \title{Thresholding tests}

\author{ Sylvain Sardy\footnote{Department of Mathematics, University of Geneva; sylvain.sardy@unige.ch}, Caroline Giacobino\footnote{Department of Mathematics, University of Geneva; caroline.giacobino@unige.ch},
Jairo Diaz-Rodriguez \footnote{Department of Mathematics, University of Geneva; Jairo.Diaz@unige.ch}}
\date{}

 \maketitle

\begin{quote}
{\bf Abstract}: We derive a new class of statistical tests for generalized linear models based on thresholding point estimators. These tests can be employed whether the model includes more parameters than observations or not. For linear models, our tests rely on pivotal statistics derived from model selection techniques. Affine lasso, a new extension of lasso, allows to unveil new tests and to develop  in the same framework parametric and nonparametric tests. Our tests for generalized linear models are based on new asymptotically pivotal statistics. A composite thresholding test attempts to achieve uniformly most power under both sparse and dense  alternatives with success. In a simulation, we compare the level and power of these tests under sparse and dense alternative hypotheses. The thresholding tests have a better control of the nominal level and higher power than existing tests.

\end{quote}





\section{Introduction}
\label{sct:intro}

Consider linear models 
\begin{equation} \label{eq:linearmodel}
 {\bf Y} = X {\boldsymbol \beta} + {\boldsymbol \epsilon}  \quad {\rm with} \quad {\boldsymbol \epsilon} \sim {\rm N}({\bf 0},\sigma^2 I_N),
\end{equation}
where ${\bf Y}$ is the $N \times 1$ response vector and $X$ is an  $N \times P$ matrix of covariates with corresponding coefficients ${\boldsymbol \beta}$.
The number $P$ of coefficients of interest can be larger than the sample size $N$.
Linear regression, linear inverse problems and analysis of variance  are examples of such models,
where $X$ is either random (e.g., regression with random covariates), or fixed (e.g., discretized basis functions like splines or wavelets, and analysis of variance).
In regression, the goal is to fit the model
using for instance the prediction risk $E\|X^{\rm new}( {\boldsymbol \beta}- \hat {\boldsymbol \beta})  \|_2^2$ for goodness-of-fit criterion.
In linear inverse problems, the goal is good estimation of ${\boldsymbol \beta}$, for instance measured
by the risk $E\| {\boldsymbol \beta}- \hat {\boldsymbol \beta}  \|_2^2$.
The primary goal of this paper is testing ${\boldsymbol \beta}$ based on thresholding techniques used in regression and inverse problems.
We are interested in testing linear null hypotheses of the form
\begin{equation} \label{eq:H0}
 H_0:\  A {\boldsymbol \beta}={\bf c},
\end{equation}
for some $R\times P$ full row rank matrix of interest~$A$.
For instance, one often desires to test a subset of ${\boldsymbol \beta}$ equal to ${\bf c}$, say
\begin{equation} \label{eq:H00}
 H_0:\  (\beta_{j_0+1}, \ldots, \beta_{P})={\bf c},
\end{equation}
for some small $j_0<N$, which amounts to
$A=[O \ I_{P-j_0}]$
where $O$ is the $(P-j_0) \times j_0$ zero matrix and $I_{P-j_0}$ is the identity matrix.
The matrix $A$ can also correspond to contrasts in analysis of variance.
In linear models with $P<N$, Fisher's $F$-test is  widely applied and based on the statistic
\begin{equation} \label{eq:Ftest}
  \frac{({\rm RSS}_{H_0}-{\rm RSS})/R}{{\rm RSS}/(N-P)} \sim F_{R,N-P}
\end{equation}
that is pivotal under $H_0$,
where ${\rm RSS}_{H_0}$ and ${\rm RSS}$ are the residual sum of squares under the null model and the full models, respectively.
We contend that one drawback of the $F$-test 
is that it is based on an indirect measure of the coefficients~${\boldsymbol \beta}$ through
the predictive measure of ${\bf Y}$ that is RSS.
\citet{arias-castro2011} show the $F$-test is suboptimal and sometimes powerless
when testing against a sparse alternative, that is, when only a few coefficients are different from zero.
A test based on a direct measure of the coefficients shall bring more power, as we will see with thresholding tests. 
Another drawback is that the $F$-test requires $P< N$ for the second degree of freedom to be positive and for the rank of $X$ to be smaller than the length of the response vector~${\bf Y}$, otherwise the estimation of variance (the denominator in~\eqref{eq:Ftest}) gives zero.

In generalized linear models \citep{NW72},  testing $H_0: {\boldsymbol \beta}={\bf 0}$ with $P\geq N$ parameters is also difficult because the model is saturated when $P\geq N$.
In the standard setting with $P<N$ fixed, letting $L({\boldsymbol \beta})$ be the likelihood function, the likelihood ratio or deviance test relies on the asymptotic distribution  
\begin{equation}\label{eq:LRS}
-2 \log \frac{\sup_{\boldsymbol \beta} L({\boldsymbol \beta})  }{L({\bf 0})} \rightarrow_d \chi^2_P
\end{equation}
under $H_0$ as $N$ tends to infinity, provided the model satisfies the conditions for asymptotic normality of maximum likelihood estimation \citep{Wilks1938}.
But asymptotic convergence is slow when $P$ is large and fails in high-dimension $P\geq N$, which motivated
\citet{GvHF11,RSSB:RSSB12152,arXiv:1706.01191v1} to propose tests based on other asymptotic distributions.
In the Gaussian case, the $F$ distribution of~\eqref{eq:Ftest} converges to the $\chi^2_P$ distribution  when $N$ gets large for a fixed $R=P$.

The  situation $P\geq N$ is difficult in testing but is well addressed  in  model selection. This paper exploits the ability of model selection methods to cope with $P\geq N$
to provide new solutions to testing.
A famous example of model selection method is lasso \citep{Tibs:regr:1996} which calculates
\begin{equation} \label{eq:lasso}
 \hat {\boldsymbol \beta}_{\lambda} \in \argmin_{{\boldsymbol \beta}\in {\mathbb R}^P} \frac{1}{2}\|{\bf Y}-  X {\boldsymbol \beta} \|_2^2 + \lambda
 \|A {\boldsymbol \beta} \|_1
\end{equation}
for a given $\lambda>0$, where $A=[O \ I_{P-j_0}]$ is the matrix that allows to not  penalize the first $j_0$ coefficients in ${\boldsymbol \beta}$.
The extension of lasso to generalized linear models replaces the quadratic loss by the negative log-likelihood \citep{ParkHastie07}.
Both estimators can be employed  whether $P<N$ or not, and are model selection techniques in the sense that the solution $\hat {\boldsymbol \beta}_{\lambda}$ in~\eqref{eq:lasso} is sparse. 
Based on these two properties we develop new tests  that continue to hold when $P\geq N$, and that have good level and power properties. 

The article proposes tests based on new pivotal statistics for Gaussian linear models and asymptotic pivotal statistics for generalized linear models, 
and is organized as follows.  
First Section~\ref{subsct:illustrexample} starts with a simple example, 
Section~\ref{subsct:general} presents the general approach, and Section~\ref{subsct:combine} shows how to combine several tests in a single one.
Section~\ref{sct:affinelasso} considers exact tests for Gaussian linear models.
Section~\ref{subsct:lassos} reviews existing thresholding estimators for testing~\eqref{eq:H00}.
Section~\ref{subsct:Ab-c} defines a new estimator called \emph{affine lasso} designed for testing the more general null hypothesis~\eqref{eq:H0}.
Section~\ref{subsct:connection} shows that thresholding tests lead to both parametric and nonparametric tests, retrieve existing tests and yield new tests in the same framework.
Section~\ref{subsct:oplus} defines the composite $\oplus$-test between lasso and group lasso.
Section~\ref{sct:GLM} considers asymptotic tests for generalized linear models.
Section~\ref{subsct:newasympivot} proposes a new asymptotic pivot.
To compare the new thresholding tests to existing tests, Sections~\ref{subsct:power1} and~\ref{subsct:power2}   perform power analyses in low- and high-dimensional settings
for Gaussian, binomial and Poisson data.
Section~\ref{sct:CR} proposes confidence regions dual to the proposed tests.
Section~\ref{sct:howtouseTT} concludes by giving recommendations on what test to use. 
Proofs are postponed to an appendix. The research is reproducible and codes are available in the {\tt qut} package in {\tt R}  \citep{qut}.

\section{Thresholding tests} \label{sct:definitions}


\subsection{An illustrative example} \label{subsct:illustrexample}

Suppose we desire to test the null hypothesis~\eqref{eq:H00} with ${\bf c}={\bf 0}$ for the  linear model~\eqref{eq:linearmodel} with $j_0=1$
to not test the intercept that corresponds to the first column of $X$.
To that aim, we consider lasso~\eqref{eq:lasso} with  $A=[{\bf 0} \ I_{P-1}]$.
Among numerous properties of lasso \citep{BuhlGeer11}, a particularly interesting one for testing the null hypothesis $H_0:\  A {\boldsymbol \beta}={\bf 0}$
 is that lasso thresholds all tested coefficients to zero if $\lambda$ is large enough. 
More precisely, the following equivalence holds:
\begin{equation}\label{eq:equivlasso}
[{\bf 0} \ I_{P-1}] \hat {\boldsymbol \beta}_{\lambda}={\bf 0} \Leftrightarrow \lambda \geq \| X^{\rm T} ({\bf y}- \bar y {\bf 1})\|_\infty. 
\end{equation}
Based on this  property, we propose a test of the form
\begin{equation} \label{eq:thresholdtestlassobeta=0}
  \phi({\bf y})=\left \{ \begin{array}{ll}
                          1 & {\rm if}\ [{\bf 0} \ I_{P-1}] \hat {\boldsymbol \beta}_{\lambda_\alpha} \neq {\bf 0} \\
                          0 & {\rm otherwise}
                         \end{array}
 \right . ,
    \end{equation}
where $\lambda=\lambda_\alpha$ is chosen for the test to have level~$\alpha$. We call it a \emph{thresholding test} because it is based on a thresholding estimator (see Definition~\ref{def:thresholdingestimator} below), here lasso.
Using~\eqref{eq:equivlasso}, one easily sees  that test~\eqref{eq:thresholdtestlassobeta=0} has the desired level by choosing $\lambda_\alpha=F^{-1}_{\Lambda_0}(1-\alpha)$
where $F_{\Lambda_0}$ is the distribution of $\Lambda_0=\| X^{\rm T} ({\bf Y}_0- \bar Y_0 {\bf 1}) \|_\infty$ and ${\bf Y}_0=_d{\bf Y}$ under $H_0$.
The test can also be simplified to
\begin{equation} \label{eq:thresholdtestlassobeta=0prime}
  \phi({\bf y})=\left \{ \begin{array}{ll}
                          1 & {\rm if}\ \| X^{\rm T} ({\bf y}- \bar y  {\bf 1}) \|_\infty>\lambda_\alpha  \\
                          0 & {\rm otherwise}
                         \end{array}
 \right . ,
    \end{equation}
which has the advantage of not having to compute $\hat {\boldsymbol \beta}_{\lambda_\alpha}$  solution to~\eqref{eq:lasso}.
The test-threshold $\lambda_\alpha$ can be evaluated  for instance by Monte Carlo simulation by
simulating~$M$ vectors ${\bf y}_0^{(1)} ,\ldots, {\bf y}_0^{(M)}$ from ${\bf Y}_0$ under $H_0$, 
calculating the corresponding
$\lambda^{(m)}=\| X^{\rm T} ({\bf y}_0^{(m)}- \bar y_0^{(m)} {\bf 1})\|_\infty$ for $m=1,\ldots,M$ and taking the upper $\alpha$-quantile. The larger $M$ the more precision on $\lambda_\alpha$.
Here the statistic $\| X^{\rm T} ({\bf y} - \bar y {\bf 1})\|_\infty$ is pivotal under $H_0$ with respect to the first entry of ${\boldsymbol \beta}$ (i.e., the intercept), but is not pivotal with respect to $\sigma$.
In the following we derive pivotal statistic with respect to all nuisance parameters.

Contrarily to Fisher's test, a thresholding test is not based on comparing two predictive measures of fit, one under the null and the other under the alternative hypothesis.
The thresholding test~\eqref{eq:thresholdtestlassobeta=0prime} is rather based on a measure (here the sup-norm) on the coefficient space through $X^{\rm T} {\bf y}$.
Also, Fisher's test depends on $X$ only through its size $(N,P)$ and the rank of $X$, which can be seen as an advantage (for the reason of tabulating the distribution), but also a drawback because the test is less specific to $X$.
On the contrary, the thresholding test~\eqref{eq:thresholdtestlassobeta=0prime} depends on~$X$ through the quantile $ \lambda_\alpha$ that is a function of the data matrix.

%

\subsection{General method} \label{subsct:general}

A test like~\eqref{eq:thresholdtestlassobeta=0prime} has good power properties compared to Fisher's test, as we will see in Section~\ref{subsct:power1}.
Test~\eqref{eq:thresholdtestlassobeta=0prime} is based on lasso, but other model selection techniques could be employed.
Depending on the model selection technique used, we either get back existing tests, or unveil new tests.
We give here a general method to derive a thresholding test $\phi$ for~\eqref{eq:H0}.
We first formally define a thresholding point estimator and test.

\begin{definition}\label{def:thresholdingestimator}
  Let $A$ be an $R\times P$ matrix of full row rank. We say that $\hat {\boldsymbol \xi}_\lambda=A\hat {\boldsymbol \beta}_\lambda({\bf Y})-{\bf c}$ is a \emph{thresholding estimator}, if 
  $$
  A \hat {\boldsymbol \beta}_\lambda-{\bf c}={\bf 0} \Leftrightarrow \lambda \ge \lambda_0({\bf y}),
  $$
  where $\lambda_0({\bf y})<\infty$. 
\end{definition}
The function of the data  $\lambda_0({\bf y})$ that provides the smallest threshold~$\lambda$ such that $A \hat {\boldsymbol \beta}_\lambda-{\bf c}={\bf 0}$ is the \emph{zero-thresholding function} \citep{CaroNickJairoMe2016}.

\begin{definition}\label{def:thresholdingtest}
 Assume $A\hat {\boldsymbol \beta}_\lambda({\bf Y})-{\bf c}$ is a thresholding estimator and consider testing the null hypothesis~\eqref{eq:H0} for the linear model~\eqref{eq:linearmodel}.
 A test function of the form
 \begin{equation} \label{eq:thresholdtest}
  \phi({\bf y})=\left \{ \begin{array}{ll}
                          1 & {\rm if}\ A \hat {\boldsymbol \beta}_{\lambda} -{\bf c} \neq {\bf 0}, \\
                          0 & {\rm otherwise.}
                         \end{array}
 \right .
    \end{equation}
    defines a \emph{thresholding test} with \emph{test-threshold} $\lambda$.
 \end{definition}

 The level of the thresholding test is the prescribed value $\alpha$ if the test-threshold $\lambda=\lambda_\alpha$ is chosen such that ${\mathbb P}(A\hat {\boldsymbol \beta}_{\lambda_\alpha}({\bf Y}_0) = {\bf c})\geq1-\alpha$,
 where ${\bf Y}_0 =_d {\bf Y}$   under $H_0: A{\boldsymbol \beta}={\bf c}$.
 The following proposition shows how to set the test-threshold. 

  \begin{proposition}\label{prop:1}
 Let $A\hat {\boldsymbol \beta}_\lambda({\bf Y})-{\bf c}$  be a thresholding estimator and $\lambda_0({\bf y})$ be its zero-thresholding function.
 Letting the \emph{null-thresholding statistic}
 \begin{equation} \label{eq:nullTS}
 \Lambda_0:=\lambda_0({\bf Y}_0)
 \quad {\rm with}\quad {\bf Y}_0 =_d {\bf Y}\quad {\rm under}\quad H_0: A{\boldsymbol \beta}={\bf c}, 
  \end{equation}
  then $\lambda_\alpha=F^{-1}_{\Lambda_0}(1-\alpha)$ is a test-threshold of level~$\alpha$ for the thresholding test~\eqref{eq:thresholdtest}. Moreover, the test simplifies to
   $$
  \phi({\bf y})=\left \{ \begin{array}{ll}
                          1 & {\rm if}\ \lambda_0({\bf y}) \geq \lambda_\alpha, \\
                          0 & {\rm otherwise.}
                         \end{array}
 \right .
  $$
\end{proposition}
The proof is immediate from Definition~\ref{def:thresholdingestimator}. 
If the inverse of $F_{\Lambda_0}$ does not exist, then a conservative test-threshold (probability of type I error less than~$\alpha$) can be obtained
using the generalized inverse $F_{\Lambda_0}^{-1}(p)=\inf \{\lambda \in {\mathbb R}: F_{\Lambda_0}(\lambda) \geq p \}$.

For linear models, Section~\ref{sct:affinelasso} shows that $\Lambda_0$ in~\eqref{eq:nullTS} is pivotal for some thresholding estimators.
So a test with exact level~$\alpha$ can be implemented.
For generalized linear models, Section~\ref{sct:GLM} proposes a null-thresholding statistic  $\Lambda_0$  that is asymptotically pivotal.

%
%

\subsection{Combining tests} \label{subsct:combine}

Suppose that test $\phi^{(1)}$ based on a first thresholding estimator  has level $\alpha$ and good power properties for a type of alternative hypothesis,
and that test $\phi^{(2)}$ based on a second thresholding estimator has  level $\alpha$ and good power properties for another type of alternative hypothesis.
It is reasonable to wish a single test $\phi$ that has level $\alpha$ and that is almost as powerful as the best of both tests regardless of the type of alternative hypothesis.
We propose the following way to combine both tests. Let $\lambda_0^{(i)}$ and $\lambda_\alpha^{(i)}$ be the zero-thresholding function and test-threshold of test~$\phi^{(i)}$ for $i\in \{1,2\}$.
The composite null-thresholding statistic
\begin{equation}\label{eq:combining}
\Lambda_0=\max \left ( \frac{\lambda_0^{(1)} ({\bf Y}_0)}{\lambda_\alpha^{(1)}}, \frac{\lambda_0^{(2)} ({\bf Y}_0)}{\lambda_\alpha^{(2)}} \right ).
\end{equation}
can be employed to develop a single test of level~$\alpha$.
The standardization by either $\lambda_\alpha^{(1)}$ or $\lambda_\alpha^{(2)}$ ensures both individual test statistics within~\eqref{eq:combining} possess the same rejection region $[1,\infty]$.

\section{Linear models} \label{sct:affinelasso}

\subsection{Existing thresholding estimators for ${\boldsymbol \beta}$} \label{subsct:lassos}
%

We are interested first  in testing the null hypothesis~\eqref{eq:H00}  that the last $P-j_0$ entries of  ${\boldsymbol \beta}$ are all null.
Many thresholding estimators of ${\boldsymbol \beta}$ already exist
to test~\eqref{eq:H00}  based on Proposition~\ref{prop:1}. 
Stepwise subset selection and  the Dantzig selector \citep{Cand:Tao:dant:2007}
are possible candidates.
Another possibility is to employ one of many versions of lasso:
these estimators are solution to the following penalized least squares problems:
\begin{equation}\label{eq:lassofamily}
 \hat {\boldsymbol \beta}\in \argmin_{{\boldsymbol \beta} \in {\mathbb R}^P}  \frac{1}{h\eta}\left (\|{\bf y}- X {\boldsymbol \beta} \|_\eta^\eta  \right )^h + \lambda
  \sum_{k=1}^K \|{\boldsymbol \beta}_{G_k} \|_j,
\end{equation}
given a partition $\{1,\ldots,P\}=\{1,\ldots,j_0\} \bigcup_{k=1}^K G_k$ and letting ${\boldsymbol \beta}_{G_k}=(\beta_p)_{p \in G_k}$. 
Depending on the partition and on the value of $j$, these estimators assume different a priori sparsity structures on ${\boldsymbol \beta}$, which leads to tests with different power properties, as we will see.
The choice of $h$ and $\eta$ also leads to different tests.
The thresholding estimators indexed by $(h,\eta,j)$ have been baptized square-root lasso  $(1/2,2,1)$ \citep{BCW11}, group square-root lasso $(1/2,2,2)$ \citep{BLS14}, LAD lasso $(1,1,1)$ \citep{CIS-215377}, lasso for $(1,2,1)$ \citep{Tibs:regr:1996} and group lasso  $(1,2,2)$ \citep{Yuan:Lin:mode:2006}. 
The partitioning of the coefficient vector ${\boldsymbol \beta}$ is subjective and depends on the problem at hand.
It should be guided by the form of the believed alternative hypothesis (see Section~\ref{subsct:power1}).
It also depends on the type of variables; for an analysis of variance for instance, the variables can be grouped into main effects, interactions and random effects.

For many thresholding estimators including those defined by~\eqref{eq:lassofamily}, \citet{CaroNickJairoMe2016} provide formula of their zero-thresholding function $\lambda_0({\bf y})$, which allows to implement thresholding tests for the null hypothesis~\eqref{eq:H00} based on Proposition~\ref{prop:1}.
\subsection{Affine lasso: a new estimator and test for $A{\boldsymbol \beta}-{\bf c}$} \label{subsct:Ab-c}

With current model selection techniques, one can not test the more general null hypothesis~\eqref{eq:H0} which depends on a matrix $A$ and a vector~${\bf c}$.
To that aim we  define the more general affine lasso.
We use the notation $A^{H}$ for the rows of $A$ with indices in $H$.

\begin{definition} \label{def:affinelasso}
Given a threshold $\lambda$, parameters $h>0$ and $\eta>0$, an $R\times P$ full row-rank matrix $A$ and a partition   $\{1,\ldots,R\}=\bigcup_{l=1}^L H_l$,
\emph{affine lasso} ($j=1$) and \emph{affine group lasso} ($j=2$) estimates are defined by
\begin{equation}\label{eq:affinelasso}
 \hat {\boldsymbol \beta}\in \argmin_{{\boldsymbol \beta} \in {\mathbb R}^P}  \frac{1}{h\eta}\left (\|{\bf y}- X {\boldsymbol \beta} \|_\eta^\eta  \right )^h + \lambda \sum_{l=1}^L \|A^{H_l} {\boldsymbol \beta}-{\bf c}_{H_l}\|_j
\end{equation}
\end{definition}
Generalized lasso \citep{CIS-255162} is a particular case of affine lasso when ${\bf c}={\bf 0}$, $L=1$ and $j=1$. 
%
%

To implement a thresholding test with affine lasso according to Proposition~\ref{prop:1}, the zero-thresholding function $\lambda_0({\bf y})$ of affine lasso is needed and given in the following proposition.

\begin{proposition} \label{prop:affinelassoZTF}
Let $A$ be a full row rank matrix, denote by $K_A$ a matrix which columns form a basis for the kernel of $A$ and by $P_{XK_A}$ the projection on the range of $XK_A$,
and let $\eta=2$.
For $(h,j)=(1,2)$, the  affine group lasso estimator  of $A{\boldsymbol \beta}-{\bf c}$
admits the zero-thresholding function 
$$\lambda_0({\bf y})=\max\limits_{l=1,\ldots,L}\| \left [(AA^{\rm T})^{-1}A X^{\rm T}  {\bf r}\right ]^{H_l} \|_2,$$
which simplifies to $$\lambda_0({\bf y})=\| (AA^{\rm T})^{-1}A X^{\rm T} {\bf r} \|_\infty$$
for affine lasso with  $(h,j)=(1,1)$, where ${\bf r}=(I-P_{XK_A})\{{\bf y}-X A^{\rm T}(AA^{\rm T})^{-1}{\bf c}\}$.
The zero-thresholding functions of affine group square-root lasso and affine square-root  lasso with $h=1/2$ are obtained by dividing the zero thresholding functions above
by $\| {\bf r} \|_2$.
\end{proposition} 

The projection matrix $P_{X K_A}$ often has a closed form expression.  For instance, $P_{X K_A}$ for lasso with an unpenalized intercept  is simply the projection matrix on the column of ones. 
If ${\rm rank}(XK_A)=N$, then no thresholding test exists since $\lambda_0({\bf y})=0$ for all ${\bf y}$.
This happens for
null hypotheses of the form~\eqref{eq:H00} with $j_0=P-1$ to test whether a single coefficient is null among $P>N$ coefficients; see \citet{wasserman2009,jasa.2009.tm08647} and \citet{Javanmard2013ConfidenceIA,RSSB:RSSB12026,vandegeer2014} in that case.

\begin{lemma}\label{lemma:ALpivotal}
Assume ${\rm rank}(X K_A)<N$. Under the assumption that $A{\boldsymbol \beta}={\bf c}$, the statistic ${\bf R}_0=(I-P_{XK_A}) \{{\bf Y}_0-X A^{\rm T}(AA^{\rm T})^{-1}{\bf c}\}/\sigma$ is pivotal with respect to ${\boldsymbol \beta}$ and  $\sigma$.
\end{lemma}

From Lemma~\ref{lemma:ALpivotal}, one sees that affine square-root (group) lasso has  a null-thresholding statistic $\Lambda_0=\lambda_0({\bf Y}_0)$ that is  pivotal  under~$H_0$.
On the contrary, affine (group) lasso leads to a null-thresholding function that is not pivotal with respect to $\sigma$.
When $P<N$, it can be made pivotal by dividing by the standard estimate of $\sigma$ under the full model.
When $P\geq N$, it can be made pivotal with the standard estimate of $\sigma$ under the null model, which amounts to affine square-root (group) lasso.
With a pivotal statistic $\Lambda_0$, Proposition~\ref{prop:affinelassoZTF}  allows to readily implement the thresholding tests of Proposition~\ref{prop:1} to test $H_0: \ A {\boldsymbol \beta}={\bf c}$.

\subsection{Connection between affine lasso tests and existing tests} \label{subsct:connection}


We illustrate connections between existing tests and thresholding tests with two classical parametric and nonparametric tests.

First, the following lemma establishes  that Fisher's $F$-test is a particular case of thresholding tests, namely group lasso.
Indeed Fisher's test belongs to the class of thresholding tests for specific values of $(j,\eta, h,L)=(2,2,1,1)$ in the definition of affine lasso~\eqref{eq:affinelasso} and
a specific standardization $Q$ of the penalty.
\begin{property} \label{lemma:Fisher} Fisher's $F$-test is a thresholding test based on the zero-thresholding function of 
 \begin{equation}\label{eq:weightedaffinelasso}
\min_{{\boldsymbol \beta} \in {\mathbb R}^P}  \frac{1}{2} \|{\bf y} - X {\boldsymbol \beta} \|_2^2 + \lambda \|Q^{1/2}(A {\boldsymbol \beta}-{\bf c})\|_2.
\end{equation}
for $Q=\{A (X^{\rm T}X)^{-1} A^{\rm T}\}^{-1}$.
\end{property}
\noindent

Likewise, LAD lasso with $(j,\eta, h,L)=(1,1,1,1)$  leads to an existing test for a particular design matrix $X$: the nonparametric distribution-free sign test \citep{Fisher1925}.
Recall the distribution-free sign test assumes pairs of observations
$(U_n, V_n)_{n=1,\ldots,N}$ on $N$ subjects following the model $V_n-U_n=\beta + \epsilon_n$ with  zero median errors. To test $H_0:\ \beta=0$, the distribution-free sign test is
based on the pivotal statistic $B=\sum_{n=1}^N 1_{\{V_n>U_n\}}$. 
To see that LAD lasso test leads to the same test, the null-thresholding statistic of LAD lasso is
 $\Lambda_0=\|X^T  {\rm sign} ({\bf Y}_0) \|_\infty=|\sum_{n=1}^N {\rm sign}(V_n-U_n)|=|2B-N|$
since in that case ${\bf Y}={\bf V}-{\bf U}$ and $X$ is simply the column vector of ones.
This statistics is pivotal with respect to $\sigma$
 and is nonparametric in the sense that its distribution is unchanged for all error distributions with zero median.

\subsection{The composite $\oplus$-test between lasso and group lasso} \label{subsct:oplus}

\citet{arias-castro2011} conclude that a test based on lasso is powerful under sparse alternatives and powerless under dense alternatives, while Fisher's or group lasso tests (see Property~\ref{lemma:Fisher}) behave the other way around.
Based on Section~\ref{subsct:combine}, we propose the composite $\oplus$-test that combines the test based on lasso (``+'' character symbolizes the coordinate-wise nature of lasso)
and the test based on group lasso (``$\circ$'' character symbolizes the $\ell_2$-ball of group lasso's penalty).
The goal of this test is to be nearly as powerful as the best test between lasso's and group lasso's tests, which we investigate in Section~\ref{subsct:power1}.

\subsection{Comparative power analysis} \label{subsct:power1}

To illustrate how thresholding tests compare with classical and more contemporary tests in terms of power,
we consider the class of alternative hypotheses  
\begin{equation} \label{eq:H1}
H_1^{s,\theta}: {\boldsymbol \beta}=\theta\cdot \pi((\underbrace{\pm 1, \ldots, \pm 1}_s,\underbrace{0,\ldots,0}_{P-s}) ^{\rm T}),
\end{equation}
indexed by $s\in \{ 0,1,\ldots,P \}$ and $ \theta \in {\mathbb R}$: $s$ controls the amount of sparsity and $\theta$ controls the signal-to-noise ratio.
Here $\pi({\bf u})$ performs a random permutation of the vector ${\bf u}$.
The sign of the coefficients $\beta_p=\pm \theta$ are random and equiprobable for $p=1,\ldots,s$. 
We say that the alternative hypothesis is sparse when $s$ is small and dense when $s$ is large with respect to $P$.
We estimate by Monte Carlo simulation power functions 
as a function of the two parameters $(s,\theta)$ indexing the alternative $H_1^{s,\theta}$-hypotheses~\eqref{eq:H1}. 
Three $X$ matrices with dimension $N=100$ and $P\in \{10,40,1000\}$ are generated according to the Monte Carlo simulation of \citet{RSSB:RSSB12152}. 
We simulate data according to  linear model~\eqref{eq:linearmodel} and 
add an intercept $\beta_0=-2$. Six tests are compared: four thresholding test (lasso, group lasso, composite lasso and LAD lasso),
the test of \citet{RSSB:RSSB12152} and Fisher's $F$-test when $P<N$.

\begin{figure}[h!]
\centering 
\includegraphics[width=\textwidth]{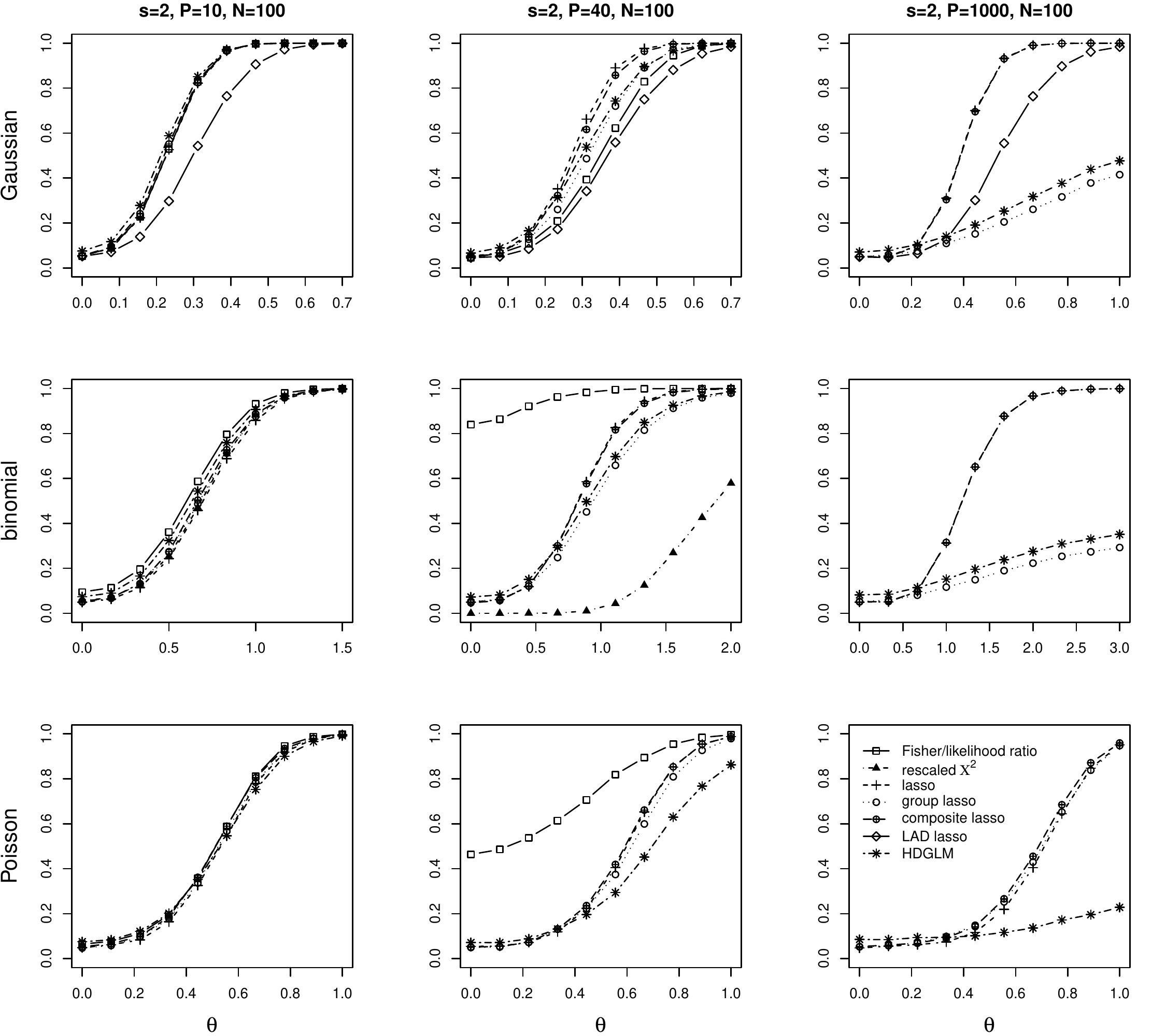}   	
\caption{Power functions estimated by Monte Carlo simulation for sparse alternative hypotheses.}
\label{fig:powersparse}
\end{figure}

\begin{figure}[h!]
\centering 
\includegraphics[width=\textwidth]{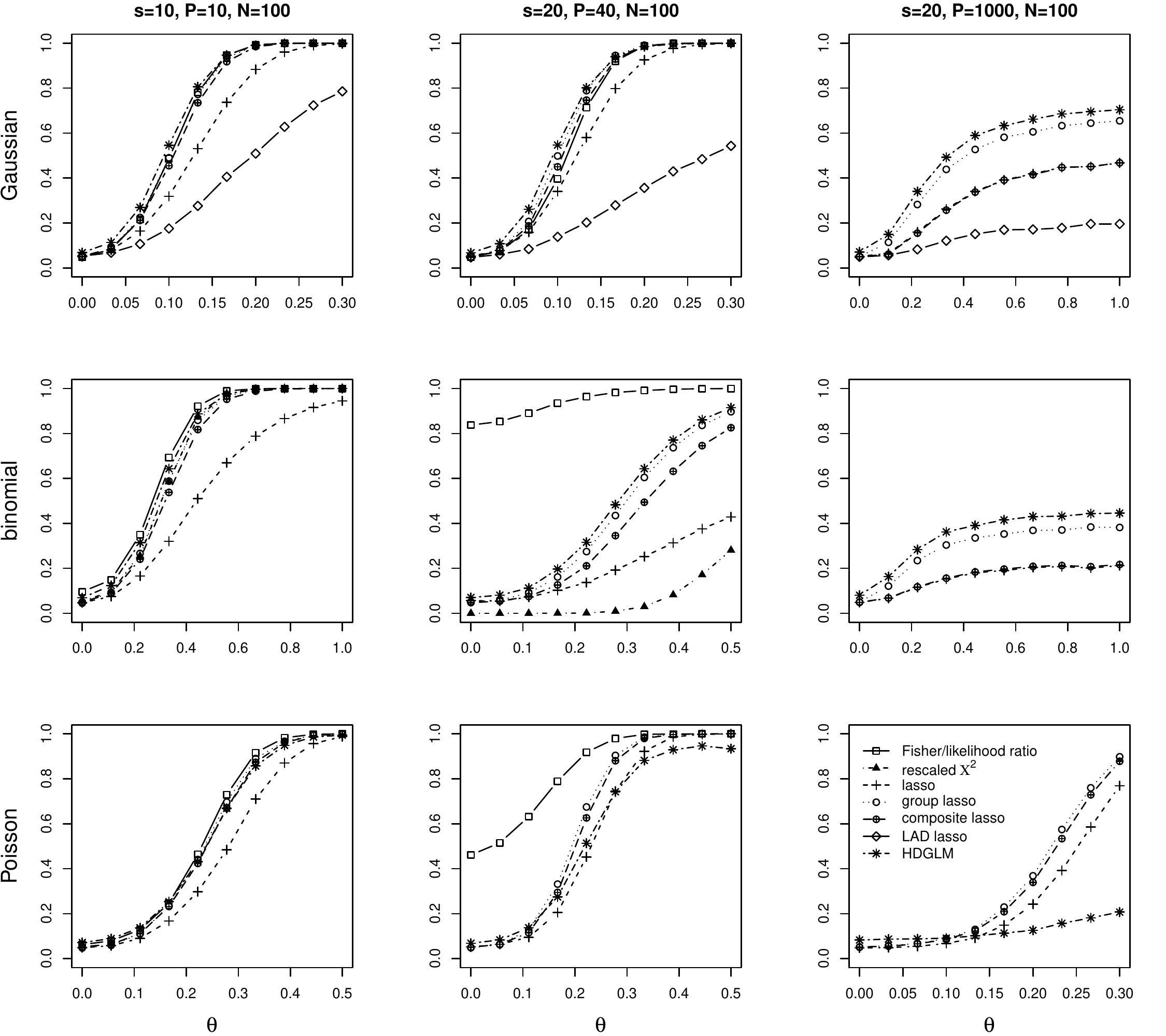}   	
\caption{Power functions estimated by Monte Carlo simulation  for dense alternative hypotheses.}
\label{fig:powerdense}
\end{figure}

The first row of Figures~\ref{fig:powersparse} and~\ref{fig:powerdense}  plot the power functions for sparse and dense alternative hypotheses, respectively
(the other rows correspond to the binomial and Poisson simulations of Section~\ref{sct:GLM}).
Interesting behaviors can be observed.
 First, comparing lasso and group lasso tests on sparse and dense situations, one sees that lasso is more powerful
 when the alternative hypothesis is sparse; on the contrary, when the alternative is dense, group lasso is more powerful. This corroborates the results of \citet{arias-castro2011}.
Second, the composite $\oplus$-test of Section~\ref{subsct:oplus}
 has power close to the most powerful test between lasso and group lasso.
Third, the nonparametric LAD lasso is the least powerful procedure, as expected; being nonparametric, its level would be robust to heavier tails than Gaussian (not shown here).
The test of \citet{RSSB:RSSB12152} (HDGLM in the plot)  is slightly off in terms of level and its power is not as good as that of the $\oplus$-test except in the dense case when $P=1000$.
Fisher's test, like group lasso's test, is better in the dense case than the sparse case. Overall the  composite $\oplus$-test is best in terms of power and in respecting the nominal level.

\section{Generalized linear models} \label{sct:GLM}

\subsection{Asymptotic pivotal thresholding statistic} \label{subsct:newasympivot}

We assume each component of the response ${\bf Y}$ has a distribution in the exponential family
\begin{equation}\label{eq:expofamily}
f_{Y_n}(y_n;\theta_n,\phi)=\exp\{(y_n\theta_n-b(\theta_n))/a(\phi)+c(y_n,\phi)\}, \quad n=1,\ldots,N,
\end{equation}
in which case the means are $\mu_n:=E[Y_n \mid {\bf x}_n]=b'(\theta_n)$. 
We assume the relation between $\mu_n$ and the covariates is linear through a link function $g$ according to
${\boldsymbol \mu}=g^{-1}(\beta_0 {\bf 1}+X{\boldsymbol \beta})$.
To test $H_0:\ {\boldsymbol \beta}={\bf 0}$, 
we derived in Section~\ref{sct:affinelasso}
thresholding tests  based on affine lasso with pivotal test statistics of the form
$\Lambda_0=\|X^{\rm T}({\bf Y}-\bar Y {\bf 1}) \|/\hat \sigma$, where $\hat \sigma/\sigma$ is pivotal under~$H_0$.
A natural extension to generalized linear models is to consider test statistics of the form 
$\Lambda_0=\|X^{\rm T}({\bf Y}- \bar Y {\bf 1})\|/D({\bf Y})$
with a denominator $D({\bf Y})$ that makes the statistic $\Lambda_0$ asymptotically pivotal.
The aims of these new tests are a tighter control of the level of the test when $P$ is large or possibly larger than $N$,
and more power than the existing tests. Indeed, most tests are based on the likelihood ratio statistic~\eqref{eq:LRS} which asymptotic chi-squared distribution
can be a poor approximation when $P$ is large and  fails when $P$ is larger than $N$.

The following theorem leads to a new asymptotic pivot for generalized linear models. 
\begin{theorem}\label{thm:glmstatistic}
Let ${\bf Y}=(Y_1,\ldots,Y_N)$ be i.i.d.~with a distribution in the exponential family~\eqref{eq:expofamily} with 
finite variance $\xi$.
Let $X$ be an $N\times P$ random matrix of $N$ vectors of 
non-degenerate covariance $\Sigma \in {\mathbb R}^{P \times P}$.
Consider the test statistic 
\begin{equation}\label{eq:T(Y)}
T({\bf Y})= \frac{\|  X^{\rm T}({\bf Y}-\bar Y {\boldsymbol 1})\|}{\sqrt{N\hat{\xi}}}.
\end{equation}
Assuming $\hat {\xi} \rightarrow_p \xi$, then
$T({\bf Y})  \rightarrow_d \left\Vert {\bf W} \right\Vert$,
where ${\bf W}\sim{\rm N}\left(0,\Sigma\right)$.
\end{theorem}
This theorem implies that $T({\bf Y})$ is asymptotically pivotal under the null hypothesis, so a test based on $T({\bf Y})$ can be employed
and a critical value asymptotically independent of the nuisance parameter $\beta_0$ can be obtained by Monte Carlo simulation as discussed in Section~\ref{subsct:illustrexample}.
For the Poisson distribution $\hat \xi=\bar Y$ is a consistent estimate of the variance under the null; likewise with $\hat \xi=\bar Y (1-\bar Y)$ 
 for the Bernoulli distribution. Section~\ref{subsct:power2} shows that the test has a good level even for finite $N$ and large $P$,
 and that the test has high power also when $P$ is larger than $N$.
 
 The statistic $T({\bf Y})$ is the zero-thresholding function $\lambda_0({\bf Y})$ of lasso for generalized linear models \citep{ParkHastie07} for certain link functions.
 When employing the canonical link, \citet{CaroNickJairoMe2016} show that the zero-thresholing function of the estimator is the numerator of $T({\bf Y})$ in~\eqref{eq:T(Y)}, which is not asymptotically pivotal.
 The following theorem states a condition on the link function for lasso to have $T({\bf Y})$ as a zero-thresholding function.


\begin{theorem}\label{thm:glmpivot}
Let ${\bf Y}$ be a random vector with a distribution in the exponential family with variance function $V(\mu)$ and known $\phi$,  and let $X$ a matrix of predictors such that 
$E[{\bf Y}]=g^{-1}(\beta_0 {\bf 1}+X{\boldsymbol \beta})$, where $g$ is the link function.
If $h=g^{-1}$ satisfies that the negative log-likelihood is convex and that $\{h'(\beta_0)\}^2=V(h(\beta_0))$, then
\begin{equation}\label{eqn:glmlassopivot}
	\lambda_0({\bf Y})=\frac{\Vert X^T\left({\bf Y}-\bar{Y}{\bf 1}\right)\Vert_\infty}{\sqrt{NV(\bar{Y})a(\phi)}},
\end{equation}
is (up to a constant) the \emph{zero-thresholding statistic} of lasso for generalized linear models.
\end{theorem}

In particular, one sees that using the inverse link function $h(x)=x$, $h(x)=x^2/4$ for $x\geq 0$, and $h(x)=(\sin(x)+1)/2$ for $x \in [-\pi/2,\pi/2]$,
respectively for Gaussian, Poisson and binomial distributions, 
the zero-thresholding function of the lasso estimator of~\citet{ParkHastie07} leads to the asymptotically pivotal test statistic~\eqref{eq:T(Y)}
with $\hat\xi= \hat \sigma^2$, $\hat \xi=\bar Y$ and $\hat \xi=\bar {Y}(1-\bar{Y})$, respectively.
These new inverse links, $h^{-1}(y)=2\sqrt{y}$ for Poisson and $h^{-1}(y)=\sin^{-1}(2y-1)$ are reminiscent of Anscombe's transforms \citep{Anscombe1948}.

\subsection{Comparative power analysis} \label{subsct:power2}

\begin{figure}[h!]
\centering 
\includegraphics[width=\textwidth]{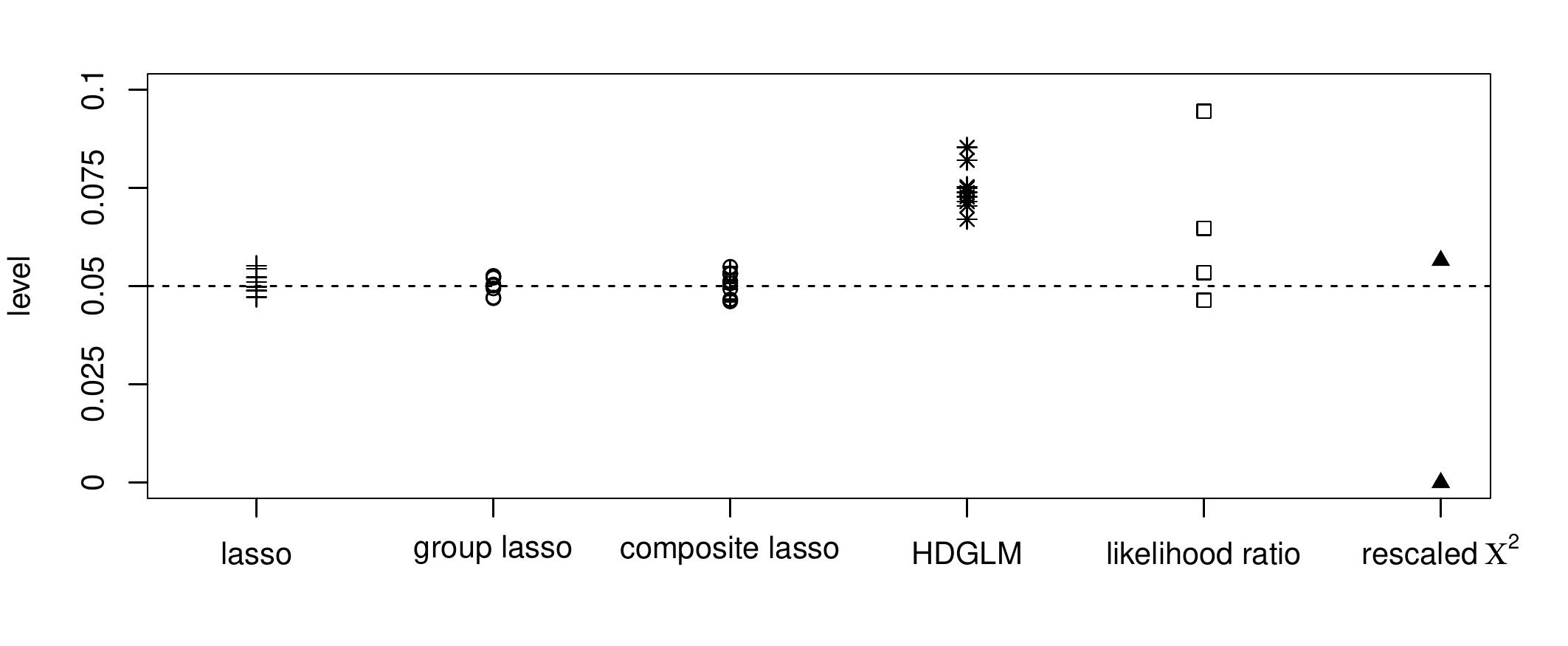}   	
\caption{Empirical levels achieved by the tests for the nine scenarios (Gaussian, binomial and Poisson and $P\in \{10,40,1000\}$). 
The values are plotted in the range $[0,0.1]$ around the nominal level $\alpha=0.05$ (dotted line).}
\label{fig:levels}
\end{figure}

We perform a Monte Carlo simulation for binomial and Poisson distributions with data generated with the canonical link based on the same setting as in Section~\ref{subsct:power2}.
Figure~\ref{fig:levels} plots the empirical levels achieved by the tests. 
 Clearly, the thresholding tests have the best control on the level. Next comes the HDGLM method of \citet{RSSB:RSSB12152} with a slight bias. Likelihood ratio test has a poor control
 of the level with two values outside the range $[0,0.1]$ of the plot (not shown here). The rescaled $\chi^2$ method of \citet{arXiv:1706.01191v1} specific to binomial performs poorly for one of the two levels.

%

Figures~\ref{fig:powersparse}  and~\ref{fig:powerdense} show the power plots for binomial (second line) and Poisson (third line).
We draw similar conclusions as in the Gaussian case, exception made for the likelihood ratio test that performs poorly due to the poor $\chi^2$ approximation when $P$ is large. Sometimes some tests appear to have higher power,
but it is important to observe that, at $\theta=0$ on the power plot, their level is larger than $\alpha=0.05$.

\section{Confidence region} \label{sct:CR}

The good power property of the thresholding tests translate to small confidence regions.
Recall that the set ${\operatorname{CR}}_{(1-\alpha)}(\boldsymbol{y})$
is  a $(1 - \alpha)$-confidence region estimator of $A\boldsymbol{\beta}$ if
\begin{equation*}
\mathbb{P} (A\boldsymbol{\beta} \in \operatorname{CR}_{(1-\alpha)}(\boldsymbol{Y})) = 1 - \alpha \quad \forall \boldsymbol{\beta} \in \mathbb{R}^P.
\end{equation*}
An example of confidence region consists of all the vectors $\boldsymbol{c}$ that are not rejected by an $\alpha$ level test of $H_0: \ A \boldsymbol{\beta} = \boldsymbol{c}$. 
Hence, a confidence region for $A\boldsymbol{\beta}$ can be defined as follows.

\begin{definition} \label{def:CR}
 Consider a thresholding test  of level~$\alpha$ for $H_0:\ A{\boldsymbol \beta}= {\bf c}$
and its test-threshold~$\lambda_ \alpha$. A $(1 - \alpha)$-confidence region estimator of $A\boldsymbol{\beta}$ is 
$$
{\operatorname{CR}}_{(1-\alpha)}(\boldsymbol{y})=\{\boldsymbol{c}\in {\mathbb R}^R:\ \lambda_{{\rm CR}}({\bf c} ; {\bf y})\leq \lambda_\alpha \},
$$
where $\lambda_{{\rm CR}}({\bf c} ; {\bf y})=\lambda_0({\bf y})$ with $\lambda_0({\bf y})$ any zero-thresholding function given in Proposition~\ref{prop:affinelassoZTF}.
\end{definition}
A possible application of confidence region is to check whether a sparse model belongs to it or not;
for instance the models along the lasso path are potential candidates.

 \section{Discussion} \label{sct:howtouseTT}
 
 Thresholding tests have good control of the nominal level of the test and  have high power for distributions in the exponential family regardless of the relative size of $N$ and $P$.
%
%
%
%
%
%
%
%

To implement a thresholding test, the choice of a thresholding estimator is required.
Like the choice of any test, the choice of a thresholding estimator is based on the data at hand and the assumed model.
We give the following recommendations. One can always use the square-root or LAD  lasso whether $N>P$ or not.
 When the alternative hypothesis is dense, one should use group lasso 
 since it has more power than lasso.
 On the contrary, when the alternative hypothesis is sparse, choosing lasso leads to more power than with group lasso's and Fisher's tests.
 When no a priori dense or sparsity assumption can be made on the likely alternative hypothesis, the composite lasso $\oplus$-test should be used. 
 When the error distribution of the data is additive but with possible outliers,  use the nonparametric LAD lasso. 
 In summary, the lasso/group lasso composite~$\oplus$-test is promising and deserves to be further studied.

\section{Acknowledgements}

We thank the Swiss National Science Foundation for the financial support of the Ph.D.~thesis of Caroline Giacobino and Jairo Diaz-Rodriguez, and Pascaline Descloux and Julie Josse for their
suggestions to better organize the paper.

\appendix \label{app:proof}

\section{Proof of  Proposition~\ref{prop:affinelassoZTF}} 

 The KKT conditions of affine lasso ($j=1$) and group lasso ($j=2$) with $K=1$ group are
$$
{\bf 0} \in X^{\rm T}(X {\boldsymbol \beta} -{\bf y})+ \lambda A^{\rm T}  {\bf s}^{(j)} 
$$
where $s_r^{(1)}\in \left \{ \begin{array}{ll}
              \{-1\} &  {\rm if}\  (A {\boldsymbol \beta})_r < c_r \\
             \left [-1, 1 \right ] &  {\rm if}\  (A {\boldsymbol \beta})_r = c_r \\
             \{ +1\}  &  {\rm if}\  (A {\boldsymbol \beta})_r > c_r 
             \end{array} \right . \ {\rm for}\  r=1,\ldots,R$,\\ \quad {\rm and} \quad
${\bf s}^{(2)}\in \left \{\begin{array}{ll}
                   (A {\boldsymbol \beta} - {\bf c})/\| A {\boldsymbol \beta} - {\bf c} \|_2 & {\rm if}\ A {\boldsymbol \beta} \neq {\bf c} \\
                   {\cal B}_R^2 & {\rm if}\ A {\boldsymbol \beta} = {\bf c}
                 \end{array}
\right .$,
where ${\cal B}_R^p$ is the $\ell_p$ unit ball in ${\mathbb R}^R$.
Consequently, the two estimators have a solution ${\boldsymbol \beta}$ satisfying $A{\boldsymbol \beta}={\bf c} $ if and only if
there exists ${\bf z}=\lambda {\bf s}^{(j)}$ with $ {\bf s}^{(j)} \in {\cal B}_R^{j/(j-1)}$ such that
  \begin{equation}\label{eq:linsys}
\left ( \begin{array}{cc}
 X^{\rm T}X & A^{\rm T }\\ A & O
\end{array} \right )
\left ( \begin{array}{c}
{\boldsymbol \beta} \\ {\bf z}
\end{array} \right )
=
\left ( \begin{array}{c}
 X^{\rm T} {\bf y} \\ {\bf c}
\end{array} \right ).
\end{equation}


  The constrained least squares problem $\min_{{\boldsymbol \beta}} \| {\bf y}- X {\boldsymbol \beta} \|_2^2$ subject to  $A{\boldsymbol \beta}={\bf c}$ (i.e., the basis of Fisher's procedure)
  has the same first order optimality conditions as those given by~\eqref{eq:linsys},  where ${\bf z}$ plays the role of the Lagrange multiplier.
  Since $A$ has full row rank and the constrained least squares fit $X \hat {\boldsymbol \beta}$ is unique, the vector ${\bf z}$ is uniquely determined by $\hat {\bf z}=(A A^{\rm T})^{-1}A X^{\rm T}({\bf y}-X \hat{\boldsymbol \beta})$.
 Since  ${\bf z}=\lambda {\bf s}^{(j)}$ in~\eqref{eq:linsys} with $ {\bf s}^{(j)} \in {\cal B}_R^{j/(j-1)}$,
 then the smallest $\lambda$ satisfying the KKT conditions for affine lasso and group lasso is $\lambda_0({\bf y})=\| \hat {\bf z}\|_{j/(j-1)}$.
 One can further identify $\hat {\bf z}$.
 Let $ {\boldsymbol \beta}_{\bf c}$ denote the unique element of $({\rm ker }A)^\perp$ such that $A {\boldsymbol \beta}_{\bf c}={\bf c}$.
 Since ${\boldsymbol \beta}_{\bf c}$ is the minimum $\ell_2$-norm solution among all solutions satisfying $A{\boldsymbol \beta}={\bf c}$, then one finds that
 ${\boldsymbol \beta}_{\bf c}=A^{\rm T} (A A^{\rm T})^{-1} {\bf c}$. 
 Let now $K_A$ denote a matrix which columns form a basis for ${\rm ker } A$.
 Then one can easily show that $X\hat{\boldsymbol \beta}=X {\boldsymbol \beta}_{\bf c}+P_{X K_A}({\bf y}-X{\boldsymbol \beta}_{\bf c})$.

 The derivation of the zero-thresholding function for $K>1$ groups and for the square-root lasso and group lasso cases are  similar and omitted.

 \section{Proof of  Lemmas~\ref{lemma:ALpivotal}}
 
 Let again $ {\boldsymbol \beta}_{\bf c}$ denote the minimum $\ell_2$-norm solution among all solutions satisfying $A{\boldsymbol \beta}={\bf c}$, and  $K_A$ denote a matrix which columns form
 a basis for  ${\rm ker } A$.
 To prove the pivotal property under $H_0$, ${\boldsymbol \beta}={\boldsymbol \beta}_{\bf c}+{\boldsymbol \gamma}$
 for some ${\boldsymbol \gamma} \in {\rm ker} (A)$.
 It follows that $(I-P_{XK_A}) ({\bf Y}_0-X {\boldsymbol \beta}_{\bf c})=(I-P_{XK_A}) (X {\boldsymbol \gamma} + {\boldsymbol \epsilon})=(I-P_{XK_A}) {\boldsymbol \epsilon}$.

  \section{Proof of Property~\ref{lemma:Fisher}}
 
 When ${\rm rank}(X)=P$ and $R$ is the rank of $A^{\rm T}$, Fisher's pivotal statistic is given in~\eqref{eq:Ftest},
 ${\rm RSS}_{H_0}-{\rm RSS}=(A \hat{\boldsymbol \beta}^{\rm LS} -{\bf c})^{\rm T} (A (X^{\rm T}X)^{-1}A^{\rm T})^{-1} (A \hat{\boldsymbol \beta}^{\rm LS} -{\bf c})$
 and $S_2^2={\rm RSS}/(N-P)$ is the unbiased estimate of variance.
%
%
 On the other hand, the zero-thresholding function of~\eqref{eq:weightedaffinelasso} is $\lambda_0^2({\bf y})=\hat {\bf z}^{\rm T} \hat {\bf z}$ where the
 Lagrange multiplier is
$
\hat {\bf z}=Q^{-1/2} (A(X^{\rm T}X)^{-1}A^{\rm T})^{-1}(A \hat{\boldsymbol \beta}^{\rm LS} -{\bf c})
$.
So $\lambda_0^2({\bf y})={\rm RSS}_{H_0}-{\rm RSS}$ if and only if $Q=(A (X^{\rm T}X)^{-1}A^{\rm T})^{-1}$.
In that case, Fisher statistics is  $F={\lambda_0^2({\bf y})/S_2^2/R}$.
So letting the null-thresholding statistic be $\Lambda_0=\lambda_0({\bf Y}_0)/S_2$, Fisher's $F$-test and the thresholding test based on the estimator~\eqref{eq:weightedaffinelasso} are identical.

\section{Proof of  Theorem~\ref{thm:glmstatistic}}

Let $M={\bf 1} {\boldsymbol \mu}_X^{\rm T}$ be the matrix of size $N\times P$ with ${\boldsymbol \mu}_X=E({\bf X})$ and ${\bf X}$ is the random vector generating covariates. 
Notice that 
\begin{eqnarray}
X^{\rm T}({\bf Y}-\overline Y{\boldsymbol 1})
								   \label{xty_ybar}
								   &=&(X-M)^{\rm T}({\bf Y}-\mu{\boldsymbol 1})+(X-M)^{\rm T}(\mu{\boldsymbol 1}-\overline Y{\boldsymbol 1}).
\end{eqnarray} 
On the one hand $(X-M)^{\rm T}({\bf Y}-\mu{\boldsymbol 1})=\sum_{n=1}^N {\bf Z}_n$ with ${\bf Z}_n=(Y_n-\mu)({\bf X}_n-{\boldsymbol \mu}_X) \in R^P$, where ${\bf X}_n^{\rm T}$ is the $n$-th (random) row of the matrix of covariates $X$ for $n=1,\ldots,N$.
The first two moments are $E[{\bf Z}_n]={\bf 0}$ and ${\rm var}({\bf Z}_n)=
\xi\Sigma$.
The central limit theorem states that
$\sum_{n=1}^N {\bf Z}_n/\sqrt{N} \rightarrow_d {\rm N}\left({\bf 0},\xi\Sigma\right)$.
On the other hand $(X-M)^{\rm T}(\mu{\boldsymbol 1}-\overline Y{\boldsymbol 1})=
(\mu-\overline Y)\sum_{n=1}^N {\bf W}_n$ with ${\bf W}_n=({\bf X}_n-{\boldsymbol \mu}_X)$. 
The first two moments are $E[{\bf W}_n]={\bf 0}$ and ${\rm var}({\bf W}_n)=\Sigma$. 
Combining central limit theorem, law of large numbers and 
Slutsky's lemma, we have that
\begin{equation}\label{convergence2}
\frac{(\mu-\overline Y)(X-M)^{\rm T}{\boldsymbol 1}}{\sqrt{N}} \xrightarrow[]{p} 0.
\end{equation}
Combining \eqref{xty_ybar}, the consistency of $\hat \xi$ and \eqref{convergence2} with Slutsky's lemma leads to
$$ 
\frac{X^{\rm T}({\bf Y}-\overline Y{\boldsymbol 1})}{\sqrt{N\hat{\xi}}} \xrightarrow[]{d} {\rm N}\left({\boldsymbol 0},\Sigma\right).
$$ 
Finally, any norm being a continuous map, we have the desired result.

  \section{Proof of  Theorem~\ref{thm:glmpivot}}

Assuming $\phi$ is known and for a fixed $\lambda$, \citet{ParkHastie07} estimate $\beta_0$ and ${\boldsymbol \beta}$ by minimizing the penalized likelihood
\begin{equation}
 {\rm PL}_\lambda(\beta_0,{\boldsymbol\beta})=  - \sum_{n=1}^N \left(\frac{Y_n \theta_n - b ( \theta_n )}{a(\phi)}  \right)  + \lambda \| {\boldsymbol \beta} \|_1.
\end{equation}
 By properties of the exponential family, we have $ E(Y_n)=b'(\theta_n)=h(\beta_0+{\bf x}_n^{\rm T}{\boldsymbol\beta})$ and ${\rm var}(Y_n)=b''(\theta_n)a(\phi)$. Consequently
$$
\left \{
\begin{array}{lll}
\frac{\partial \theta_n}{\partial \beta_0}& = &\frac{h'(\beta_0+{\bf x}_n^T{\boldsymbol\beta})}{b''(\theta_n)}\\
\nabla_{{\boldsymbol \beta}} \theta_n& = &\frac{{\bf x}_n h'(\beta_0+{\bf x}_n^T{\boldsymbol \beta})}{b''(\theta_n)}.
\end{array}
\right .
$$
By assumption ${\rm PL}_\lambda$ is convex,  so the point $(\hat \beta_0, {\bf 0})$ belongs to the
minimum set of ${\rm PL}_\lambda$ if and only if ${\bf 0}$ is a subgradient of ${\rm PL}_\lambda$ at $(\beta_0, {\boldsymbol \beta})=(\hat \beta_0, {\bf 0})$.
This is equivalent to
$$
\left \{
\begin{array}{lll}
\sum_{n=1}^N \left(\frac{h'(\beta_0)}{b''(\theta_n)a(\phi)}\left(y_n-h(\beta_0)\right)\right)&=&0\\ 
\sum_{n=1}^N \left(\frac{{\bf x}_n h'(\beta_0)}{b''(\theta_n)a(\phi)}\left(y_n-h(\beta_0)\right)\right)+\lambda [-1,1]^P &\ni& {\bf 0}.
\end{array}
\right .
$$
A solution exists if and only if $h(\beta_0)=\bar{y}$ and $\lambda$ at least as large as
$\lambda_0({\bf y})=\left\Vert\frac{h'(\beta_0)X^T({\bf y}-\bar{y}{\bf 1})}{V(h(\beta_0))a(\phi)}\right\Vert_\infty$, 
where $V$ is the variance function such that $V(h(\beta_0))=b''(\theta)$.
%
So if $|h'(\beta_0)|=\sqrt{V(h(\beta_0))}$, we obtain  the desired zero-thresholding function $\lambda_0({\bf y})$ up to the constant $\sqrt{N a(\phi)}$.

\bibliographystyle{plainnat}
\bibliography{article_bis}

\end{document}